\documentclass[
 reprint,
 nofootinbib,
 showkeys,
 amsmath,
 amssymb,
 aps,
 pra,
 floatfix,
 superscriptaddress,
 longbibliography
 ]{revtex4-2}

\usepackage{graphicx}
\usepackage{multirow}
\usepackage[dvipsnames]{xcolor}
\usepackage{dcolumn}
\usepackage{bm}
\usepackage{dsfont}
\usepackage{physics}
\usepackage{hyperref}

\usepackage{ulem}

\begin{document}

\author{I.~V.~Iorsh}
\email{Contact author: ivan.iorsh@queensu.ca}
\affiliation{Department of Physics, Engineering Physics and Astronomy, Queen’s University, Kingston, Ontario K7L 3N6, Canada}
\author{L.~Paroski}

\affiliation{Department of Physics, Engineering Physics and Astronomy, Queen’s University, Kingston, Ontario K7L 3N6, Canada}

\title{Long range spatial correlations in the periodically driven transverse field Ising model.}

\date{\today}

\begin{abstract}
We consider a periodically driven transverse field Ising model and study the long-time behavior of the correlations between the excitations in the spin chain. We show that the longest correlation length is defined by the interference between the topological defects induced by the analog of the Kibble-Zurek mechanism and those induced by the Floquet resonance. We show that although the correlations always have a finite correlation length since the system is integrable, the correlation length can be made arbitrarily large by tuning the drive period. 
\end{abstract}

\maketitle

\section{Introduction}
The last decade has witnessed substantial progress in our understanding of the evolution of periodically driven quantum systems thanks to both theoretical~\cite{Dziarmaga2010,Polkovnikov2011,Bukov2015,DAlessio2016,Dutta2015,Calabrese2016} and experimental advancements~\cite{Greiner2002,Simon2011,Bakr2010,Bernien2017,Islam2015}. Experimental setups include but are not limited to ultracold atoms~\cite{gross2017quantum} and ions~\cite{PhysRevLett.123.213605} in optical lattices, semiconductor quantum dots~\cite{xu2007coherent} and superconducting qubits~\cite{nguyen2024programmable}. It was soon recognized that periodically driven quantum systems can be described by effective static Hamiltonians in the regime where the driving frequency exceeds characteristic energy scales of the system. This triggered the theoretical advancements in Floquet engineering and the realization of new effective models via driving~\cite{PhysRevX.4.031027}. Periodically driven quantum systems reveal a number of phenomena not present in equilibrium systems such as dynamical phase transitions~\cite{PhysRevLett.110.135704,Heyl2018}, Floquet topological transitions~\cite{Oka2009,Kitagawa2010,Lindner2011,Kitagawa2011,Mukherjee2018,Nathan2015,Rudner2013}, many-body localization~\cite{PhysRevLett.114.140401,PhysRevResearch.5.013094,PhysRevB.107.115132,PhysRevLett.124.190601}, and time crystals~\cite{RevModPhys.95.031001}.

The integrability of quantum systems plays a crucial role in the long-time evolution of the periodically driven closed quantum system. While the non-integrable system eventually heats up to infinite temperature (which may be preceded by the exponentially long pre-thermalization phase~\cite{mori2018thermalization}), the integrable systems at late times can be characterized by the generalized periodic Gibbs ensemble~\cite{PhysRevLett.112.150401} with infinitely many conserved quantities. At the same time, it was shown that in a certain class of integrable systems which can be mapped to the non-interacting fermions, novel approximate conservation laws emerge in addition to those dictated by the generalized Gibbs ensemble~\cite{PhysRevB.82.172402}. The seminal example is the periodically driven transverse-field Ising model described by the Hamiltonian:
\begin{align}
    H(t) = -J \sum_i s_i^x s_{i+1}^x - g(t) \sum_i s_i^z.
\end{align}
\begin{figure}[h!]
    \centering
    \includegraphics[width=1\linewidth]{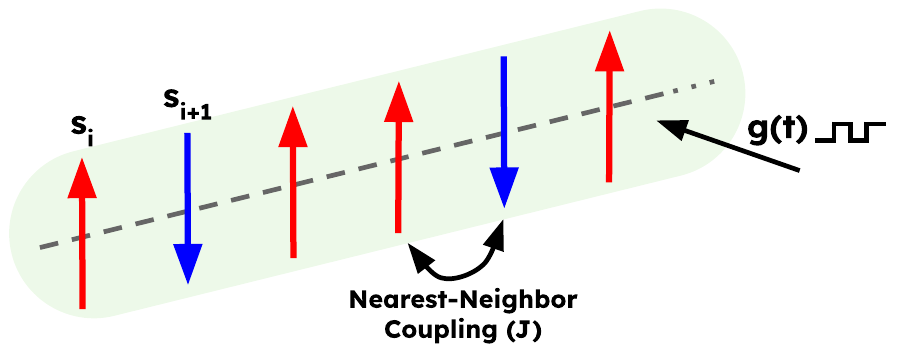}
    \caption{Schematic of the 1D transverse-field Ising chain. Spins $s_i$ (red and blue arrows) denote up and down configurations that interact via the nearest-neighbor coupling $J$. The entire chain is subject to a uniform, time-dependent transverse field $g(t)$ following a periodic rectangular drive. 
    }
    \label{fig:cond_ising}
\end{figure}
In~\cite{PhysRevB.82.172402} it has been shown that for the simplest profile of the time periodic magnetic field $g(t)=g_0\cos(\omega t)$ and for amplitude and frequency of the drive satisfying $J_0(2g_0/\omega)=0$, where $J_0$ is the Bessel function, the average magnetization $m_z =\langle s_{i}^{z}\rangle$ becomes a conserved quantity~\cite{das2010exotic} not described by the generalized Gibbs ensemble. Similar freezing behavior was later demonstrated for other driving protocols~\cite{bhattacharyya2012transverse,das2025insights}. Therefore, late-time behavior of even integrable systems can be non-trivial, demonstrating emergent conservation laws.
In this paper, we consider the periodically driven transverse field Ising model (TFIM), shown schematically in Fig.~\ref{fig:cond_ising}. 

It is known that the static TFIM experiences a quantum phase transition at $g_0=g_{cr}=J$ when the gap in the spectrum closes. Moreover, in the quench set-up when the magnetic field changes linearly and passes $g_{cr}$, the topological defects, domain walls, are excited in the system according to the Kibble-Zurek mechanism~\cite{PhysRevA.111.042207}. The late-time behavior of the system will be governed by the statistical properties of these generated defects. In our work, we focus on the correlation functions of the excitations of the Ising chain and show that they may exhibit very slow spatial decay in the vicinity of the resonances.

The paper is organized as follows. In Section II, we provide the details of the Jordan-Wigner transformation allowing one to map the TFIM to an ensemble of periodically driven non-interacting two-level systems, derive the general expressions for the correlation functions of the stroboscopic averages of the excitations, and show how the slow decay emerges at the Floquet resonances. Finally, the conclusions of the article are presented in Section III. 

\section{Model}
\subsection{Transverse-field Ising model and Bogoliubov representation}

We consider the 1D transverse-field Ising model 
\begin{align}
    H(t) = -\sum_i s_i^x s_{i+1}^x - g(t) \sum_i s_i^z,
\end{align}
subject to a periodic rectangular drive
\begin{align}
    g(t) &= \begin{cases} g_0~, & 0 \le t <  \alpha T~ \\ g_1~, &  \alpha T \le t < T~ \end{cases} \label{eq:drive}
\end{align}

with $g_0<1, g_1>1$, $\alpha<1$. The Hamiltonian is normalized to the exchange constant $J$. The case $g=1$ corresponds to a quantum critical point where the gap closes, and the energy dispersion is given by $\varepsilon(k,g) = \sqrt{g^2 - 2g\cos k + 1}$.

The Jordan--Wigner transformation maps the spin operators to spinless fermions $c_i$,
\begin{align}
    s_i^z &= 1 - 2\, c_i^\dagger c_i~, \\
    s_i^x &= \prod_{j<i}\!\big(1 - 2\, c_j^\dagger c_j\big)\,\big(c_i + c_i^\dagger\big)~.
\end{align}

A subsequent Fourier transform
\begin{align}
    c_j = \frac{1}{\sqrt{N}}\sum_k e^{ikj}\, c_k~,
\end{align}
with periodic boundary conditions and $J=1$, brings the Hamiltonian to the quadratic form
\begin{align}
    H \;=\; \sum_k \Big[\, (g(t) - \cos k)\;c_k^\dagger c_k \;+\; \sin k\;\big( c_k^\dagger c_{-k}^\dagger + c_{-k}\,c_k \big)\,\Big]~,
\end{align}

which decouples into independent $(k,-k)$ sectors. In Nambu spinor notation $\psi_k = (c_k,\, c_{-k}^\dagger)^T$, each sector is a two-level system, 
\begin{align}
    H_k = \psi_k^\dagger \big[(g(t)-\cos k)\,\tau_z + \sin k\,\tau_x\big]\psi_k~=\psi_k^{\dagger}\mathcal{H}_k\psi_k, \label{eq:Hk}
\end{align}
with $\tau_{x,z}$ the Pauli matrices. (The full Hamiltonian is $H = \sum_{k>0} H_k$, summed over independent pairs.) At the initial time $t=0^-$, the eigenstates of the Hamiltonian are given by:
\begin{align}
    |\!\downarrow_k\rangle = \begin{pmatrix} \sin\theta_k \\ -\cos\theta_k \end{pmatrix}~, \qquad
    |\!\uparrow_k\rangle = \begin{pmatrix} \cos\theta_k \\ \sin\theta_k \end{pmatrix}~,
\end{align}
with the Bogoliubov angle defined by
\begin{align}
    \tan(2\theta_k) = \frac{\sin k}{g_0 - \cos k}~. \label{eq:bogoangle}
\end{align}
We are interested in the evolution of the system at the stroboscopic times $t_N=NT$. At these times, the spinors at each $k$ can be evaluated by multiplying them by an $N$-th power of the evolution operator over the period $U_T$:
\begin{align}
U_{k,T} = e^{-i(1-\alpha)T\mathcal{H}_k(g_1)}\, e^{-i\alpha T\mathcal{H}_k(g_0)}.
\end{align}
At stroboscopic times, the many-body state vector is given by
\begin{align}
    |\Psi(t)\rangle = \prod_{k>0}\!\Big(\alpha_k(t) + \beta_k(t)\,\gamma_k^\dagger \gamma_{-k}^\dagger\Big)|0\rangle~,
\end{align}
where $|0\rangle$ is the Bogoliubov vacuum, $\gamma_k^{\dagger}$ are the creation operators of the Bogoliubov excitations and where $\alpha_k(t) = \langle\downarrow_k|U(t)|\downarrow_k\rangle$ and $\beta_k(t) = \langle\uparrow_k|U(t)|\downarrow_k\rangle$ are the amplitudes for mode $k$ to remain in its ground state or to be excited. The average defect density is $n(t) = L^{-1}\sum_k |\beta_k(t)|^2$, where $L$ is the chain length.
The spatial correlations between excitations are given by:
\begin{align}
    \rho(r,t) = \frac{1}{L}\sum_i \langle\Psi(t)|\,n_i n_{i+r}\,|\Psi(t)\rangle - n(t)^2~.
\end{align}

Expanding the four-fermion expectation value yields, for $r > 0$,

\begin{align}
    \rho(r,t) = \Bigg|\frac{1}{L}\sum_k \alpha_k^*(t)\beta_k(t)\, e^{ikr}\Bigg|^2 - \Bigg|\frac{1}{L}\sum_k |\beta_k(t)|^2\, e^{ikr}\Bigg|^2~. \label{eq:rho}
\end{align}
Taking the thermodynamic limit we substitute $L^{-1}\sum_k \rightarrow \int (2\pi)^{-1}dk$.
The amplitudes $\beta_k,\alpha_k$ can be found from the product of exponentials of $2\times 2$ matrices and yield:
\begin{align}
&|\beta_k(NT)|^2=\frac{\sin^2 k(g_0-g_1)^2\sin^2(\epsilon_1(1-\alpha)T)}{\epsilon_0^2\epsilon_1^2\sin^2\Theta}\sin^2(N\Theta), \label{eq:betsq}\\
&\cos\Theta=\cos\left(\epsilon_0\alpha T\right)\cos\left(\epsilon_1(1-\alpha)T\right)\nonumber\\&-\frac{1+g_0g_1-(g_0+g_1)\cos k}{\epsilon_0\epsilon_1}\sin\left(\epsilon_0\alpha T\right)\sin\left(\epsilon_1(1-\alpha)T\right)
\end{align}
If we average over later stroboscopic times, the term $\sin^2(N\Theta)$ in Eq.~\eqref{eq:betsq} vanishes, leaving only the prefactor $1/2$. Thus, the time-averaged quantities yield:
\begin{align}
&|\beta_k|^2 =\frac{1}{2}\frac{\sin^2 k(g_0-g_1)^2\sin^2(\epsilon_1(1-\alpha)T)}{\epsilon_0^2\epsilon_1^2\sin^2\Theta},\\
&\alpha_k^*\beta_k=-\frac{e^{i\epsilon_0T/2}\sin k(g_0-g_1)}{2\epsilon_0\epsilon_1\sin^2\Theta}\cos\left(\epsilon_0\alpha T\right)\sin\left(\epsilon_1(1-\alpha)T\right)\nonumber\\&\times\left[1+\zeta \tan\left(\epsilon_0\alpha T\right)\cot\left(\epsilon_1(1-\alpha)T\right)\right],
\end{align}
where $\zeta= (1+g_0g_1-\cos k(g_0+g_1))/(\epsilon_0\epsilon_1)$.
To find the correlation function, we then need to evaluate the integrals over the Brillouin zone. 

Upon substitution $z=e^{ik}$, the integral over the Brillouin zone transforms to the integral over the unit circle. The integral at large distances $r\gg 1$ is dominated by the residues of the poles inside the unit circle that are closest to the unit circle.

\subsection{Diverging correlation length in the vicinity of the freezing point}
In what follows we will use the following protocol: $g_0=g<1$, $g_1=1/g_0 >1$, $\alpha =(1+g_0)^{-1}$. In this case, $\epsilon_0=\sqrt{g_0}\,\epsilon, \quad \epsilon_1=\epsilon/\sqrt{g_0}$, where $\epsilon^2= (g_0 + g_0^{-1})-2\cos k$, and moreover $\epsilon_0\alpha T =\epsilon_1(1-\alpha )T=\epsilon \tilde{T}$, where $\tilde{T}= T\sqrt{g_0}/(1+g_0)$. With such a parametrization the expressions for $|\beta_k|^2$ and $\alpha_k^*\beta_k$ can be substantially simplified yielding:
\begin{align}
&|\beta_k|^2 = \frac{1}{2} \frac{1}{1+\frac{(g+1)^2}{(g-1)^2}\tan^2(k/2)\cos^2(\epsilon \tilde{T})},\nonumber \\
&\alpha^*_k\beta_k  = \frac{e^{i\epsilon \tilde{T}}}{2} \frac{\frac{(g+1)}{(g-1)}\tan(k/2)\cos(\epsilon \tilde{T})}{1+\frac{(g+1)^2}{(g-1)^2}\tan^2(k/2)\cos^2(\epsilon \tilde{T})}. ~\label{Eq:forms}
\end{align}
In order to take the corresponding integrals, we introduce $z=e^{ik}$ and take the integral over the unit circle $|z|=1$. At large distances $r$, it is expected that the correlation function will be dominated by the residue of the poles closest to the unit circle. However, if the pole $z^*$ lies on the real axis, then the residue contribution of this leading pole for $|\beta_k|^2$ and $\alpha_k^*\beta_k$ differs only by a phase factor, and therefore the sole pole contribution cancels for the correlation function. Therefore, one either needs to consider the interference of the two poles' contributions or to look for the zeros of the denominator away from the real axis. It can be shown that all of the zeros of the denominator are situated in the vicinity of the real axis. Therefore, the only possible non-vanishing contribution for the correlation function at large distances may come from the interference of two poles.  Let us write down the denominator $D(z)$ explicitly expressed via $z=e^{ik}$:
\begin{align}
D(z) = 1-\left(\frac{(g+1)(z-1)}{(g-1)(z+1)}\right)^2 \cos^2\left(\tilde{T}\sqrt{g+\frac{1}{g}-z-\frac{1}{z}}\right).
\end{align}
One zero of the denominator, $z=z_+=g$ can be identified immediately. It can be shown that this is the only zero for real positive $z<1$. If $g\approx 1-\delta g,\quad \delta g\ll 1$ then the derivative of the denominator at $z=z_+$ equals 
\begin{align}
D'(z)|_{z=z_+}\approx \frac{2}{\delta g}.
\end{align}
The numerator is equal to 1 and the phase factor $e^{i\epsilon \tilde{T}}$ is equal to 1, because the energy $\epsilon=0$. We note that this point can be regarded as the point where the energy gap closes since $\epsilon=0$, which happens at complex $k=\arccos((g+g^{-1})/2)$. This pole can be viewed as corresponding to the domain walls originating due to the periodic quenches of the system through the quantum phase transitions. Since the residue for this pole is the same for $|\beta_k|^2$ and $\alpha_k^*\beta_k$, the total correlation function vanishes if we account only for this pole.

\begin{figure}[!htbp]
    \centering
    \includegraphics[width=0.9\linewidth]{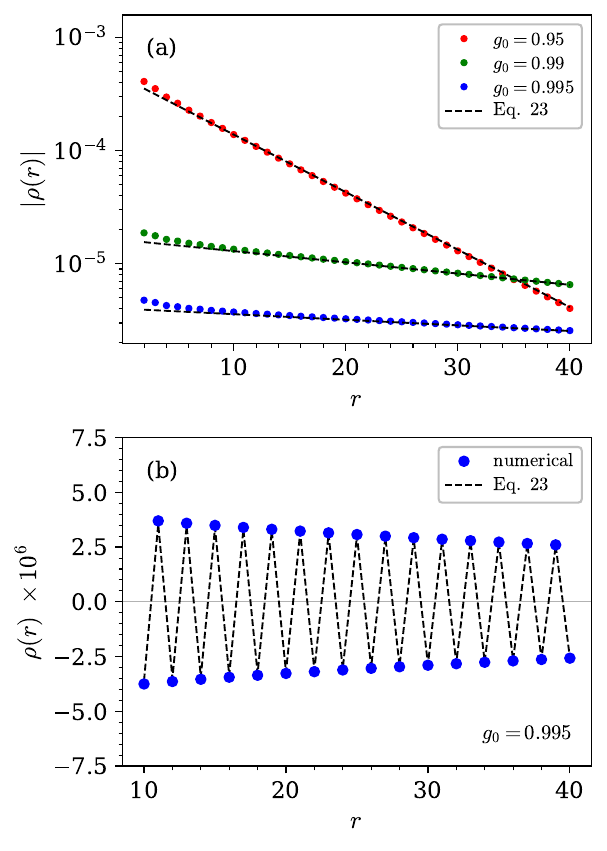}
    \caption{Correlation function $\rho(r)$ at fixed period $T = \pi/2$, compared to Eq.~\eqref{Analyt}. (a) $|\rho(r)|$ for $g_0 \in \{0.95,\, 0.99,\, 0.995\}$. (b) Signed $\rho(r)$ at $g_0 = 0.995$ over $r \in [10,\, 40]$.}
    \label{fig:2}
\end{figure}

We thus need to account for the second pole. Since there exist only negative zeros for the denominator $D(z)$ and we are looking for the poles close to the unit circle, we will expand $z$ in the vicinity of $z=-1$. The second pole in the vicinity of $z=-1$ is only present for specific values of $\tilde{T}$, specifically when $\cos(\epsilon(z)|_{z=-1}\tilde{T})=0$, i.e. when 
\begin{align}
\tilde{T} = \frac{\left(\frac{\pi}{2}+\pi m\right)}{\sqrt{g+1/g+2}}
\end{align}
For this $\tilde{T}$, the pole is given by:
$z_-=-1+4\delta g/\pi$. The residue at this pole is given by $D'(z)|_{z=z_-}=-\pi/(2\delta g)$, the numerator is -1 and the phase factor is $e^{i\pi/2 (1+\delta g^2/\pi)}$. We note that, unlike the first pole, $z=z_-$ only approaches the unit circle and thus decays slowly only for the specific values of the driving period $T$. This happens due to the fact that it is associated with the Floquet resonances in the periodically driven system and thus depends strongly on the drive frequency.

Calculating the integrals as the contribution of the two poles nearest to the unit circle, we obtain for the long-distance estimate for the correlation function:
\begin{align}
\rho(r)|_{r\gg 1} \approx(-1)^{|r-1|} \frac{1}{2\pi}\delta g^2 (z_+|z_-|)^{|r-1|} ~\label{Analyt}
\end{align}
In Fig.~\ref{fig:2}, we see the dependence of the correlation function on distance for different values of $\delta g$. It can be seen that Eq.~\eqref{Analyt} approximates the correlation function behavior very accurately.

We note that the origin of the interference in the correlation function becomes apparent from the rapidly oscillating phase term $(-1)^{|r-1|}$. The interference occurs between the excitations at $k\approx 0$ excited in the analogue of the Kibble-Zurek mechanism, since the gap at $k=0$ closes when $g$ passes unity, and the Floquet resonance at $k\approx \pi$ since the period is tuned in such a way that the corresponding driving frequency is in the vicinity of the resonance $\omega=2\pi/T\approx 2\epsilon(k)|_{k=\pi}$. This $\pi$ phase shift is reflected in the $(-1)^{|r-1|}=e^{i\pi |r-1|}$ phase factor.

\begin{figure}[t!]
    \centering
    \includegraphics[width=0.9\linewidth]{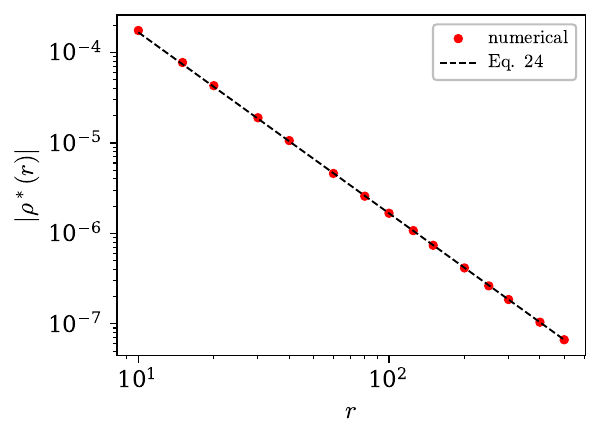}
    \caption{Optimally-tuned envelope $|\rho^\ast(r)| = \max_{\delta g}\,|\rho(r,\, \delta g)|$ at $T = \pi/2$, compared to Eq.~\eqref{rho_optimal}.}
    \label{fig:3}
\end{figure}

We note that for each specific $r\gg 1$ there exists $\delta g^* = 2\pi/(\pi r+4r) \ll 1$ which maximizes the correlation function at this distance. The correlation function at optimal $\delta g$, $\rho^*$ scales with $r$ as
\begin{align}
\rho^*  \approx  \frac{2\pi}{(\pi+4)^2 r^2}e^{-2}~\label{rho_optimal}
\end{align}
Thus, at optimal driving amplitude one can achieve $r^{-2}$ algebraic decay of the correlation function with distance. The corresponding plots are shown in Fig.~\ref{fig:3}.

Let us discuss the relevance of the obtained results to the previously discussed statistical properties in driven transverse field Ising models (TFIM). As expected for the sudden-quench protocol, the correlation length scales as $\delta g^{-1}$~\cite{de2010quench}. Specifically, the correlation length $\xi\approx(\delta g (1+4/\pi))^{-1}$. The interference origin of the correlation function and the emerging phase factor lead to the rapid sign changes of the correlations. In the presence of decoherence, the oscillations would be suppressed, leaving only the exponential suppression of the correlations. 

\section{Conclusion}
To conclude, we have considered the spatial correlations between topological defects generated in the periodically driven TFIM with a rectangular drive protocol. We have shown that by tuning the period of the oscillations to the bandwidth of the undriven system, one may induce interference between the correlations induced by the sudden quench through the quantum phase transition and the Floquet resonance. This interference induces the most slowly decaying correlations in the system, superimposed with rapid oscillations with a single-unit-cell period. The considered oscillations may be probed experimentally in a cold-atom quantum simulator with a periodically shaken optical lattice. The results highlight another example of the versatility of control over the statistical properties in condensed-matter systems subject to periodic drive.

\end{document}